# Improvement studies on neutron-gamma separation in HPGe detectors by using neural networks

Serkan AKKOYUN[1,*], Tuncay BAYRAM[2], S. Okan KARA[3]

[1] *Department of Physics, Cumhuriyet University, Sivas 58140, TURKEY*

[2] *Department of Physics, Sinop University, Sinop 57000, TURKEY*

[3] *Department of Physics, Nigde University, Nigde, TURKEY*

___

**Abstract.** The neutrons emitted in heavy-ion fusion-evaporation (HIFE) reactions together with the gamma-rays cause unwanted backgrounds in gamma-ray spectra. Especially in the nuclear reactions, where relativistic ion beams (RIBs) are used, these neutrons are serious problem. They have to be rejected in order to obtain clearer gamma-ray peaks. In this study, the radiation energy and three criteria which were previously determined for separation between neutron and gamma-rays in the HPGe detectors have been used in artificial neural network (ANN) for improving of the decomposition power. According to the preliminary results obtained from ANN method, the ratio of neutron rejection has been improved by a factor of 1.27 and the ratio of the lost in gamma-rays has been decreased by a factor of 0.50.

**KeyWords**: HPGe detectors, gamma-ray tracking, artificial neural network, Geant4, neutron-gamma separation

___

## 1. Introduction

The most common reaction type used in nuclear structure studies is heavy-ion fusion-evaporation (HIFE) reaction. In these reactions, a number of neutrons are emitted from the compound nucleus competing with the gamma-rays. Especially in exploring exotic nuclei lying around neutron drip-line by using relativistic ion beams (RIBs), the number highly increases. These neutrons cause unwanted background in the gamma-ray spectra and have to be rejected. Therefore, it is essential to separate these neutrons from the gamma-rays. The specific aim of this work is performing this separation.

In this study, the Geant4 [1] Monte Carlo simulations of highly efficient gamma-ray tracking detector AGATA [2] and interactions of the gamma-rays and neutrons within these detectors have been performed. In these detectors, neutrons interact inside the detector material via elastic or inelastic scatterings. Elastic scattering points in the detectors can be identified and rejected in the gamma-ray tracking process. But in the case of inelastic scatterings of neutrons, the gamma-rays are also emitted by the recoiling nuclei inside the detector materials. The

___

existence of the recoiling energy points in the detectors helps the determining of the three differences between neutrons and the gamma-rays. In Section 2, these three differences called as criteria will be mentioned in detail. After the simulations, by using the forward gamma-ray tracking algorithm [3], the previously determined three criteria for separation have been obtained. These criteria are as follows: energy deposited in the first interaction point, scattering angle of the radiation in the first interaction point and figure-of-merit (FM) value of the interaction point of the cluster formed in the tracking process. Finally, these three criteria and radiation energy have been used in artificial neural network (ANN) method for improving of the separation power. In the previous work [4] in which three criteria for separation have been successfully determined, in such a way that specific gates are applied on the criteria and the separation is performed. According to this method, all points inside the gates are rejected. However in the present work, ANN could identify the interaction points and could not reject them even if they are inside the gates or could reject even if they are not inside the gates. This is the one great success of the ANN. By using this method, the rejection ratio increased for neutrons and decreased for the gamma-rays by some factors.

Recently, ANNs have been successfully used in many fields including discrimination of neutrons and gamma-rays [5-8]. In Ref. [8], these three criteria have been consistently obtained by using ANNs. It can be clearly seen that the ANN method, which has speed advantage, is consistent with the experimental results.

## 2. Theoretical Framework

A brief explanations of AGATA detectors and the simulations of these detectors and interactions are given in Section 2.1. The tehniques for forward gamma-ray tracking are summarized in Section 2.2. Finally, ANN method which are widely used tool in many filds are mentioned in Section 2.3.

## 2.1 AGATA Detectors and Simulations

Two powerful detection systems based on gamma-ray tracking have been developed and under operation, AGATA in Europe [2] and GRETA in the USA [9]. In this work, only AGATA detectors have been considered and simulated. The AGATA detector array consists of 180 closed-end coaxial n-type HPGe crystals which are arranged in $4\pi$ geometry around the centered source position (Fig. 2.1). The inner radius of the array is about 22.5 cm. Each

Serkan AKKOYUN, Tuncay BAYRAM, S. Okan KARA

AGATA detector crystal is 36-fold electronically segmented and the resulting total segments provide unequalled position sensitivity that enables us to track gamma-ray traces correctly. The 1/15 of the array is currently under operation in GSI laboratory [10].

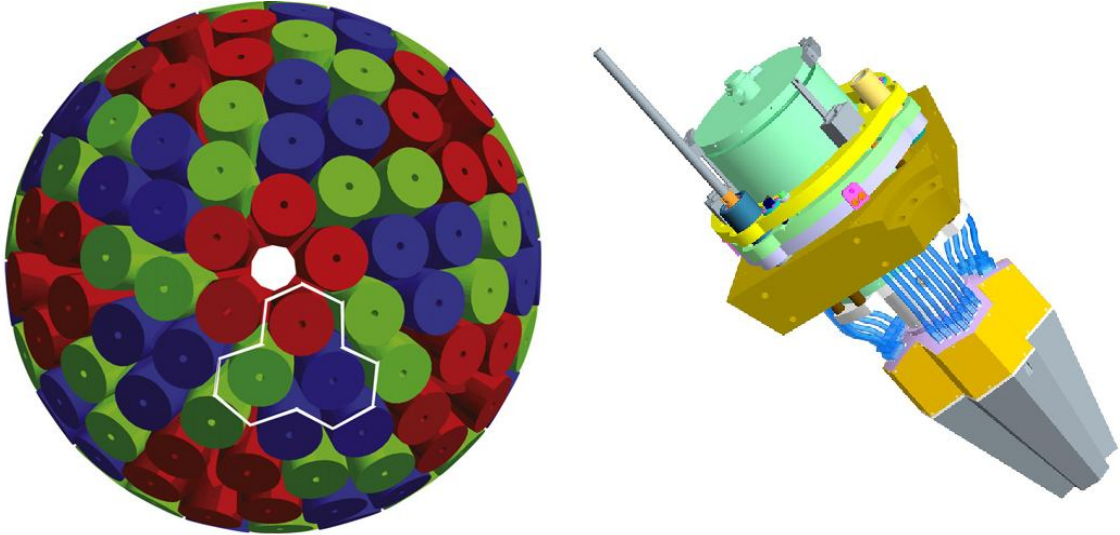

**Figure 2.1** AGATA detector array in $4\pi$ geometry (left). Each three detector are packed together using same cryostat (right) [2]

The realistic simulation of AGATA detectors and the radiation interactions inside the detectors were simulated under the C++ classes of Geant4.9.2 Monte Carlo simulation program [1]. This provide a full description of the microscopic interactions of radiation with matter, as well as tools to implement the geometry of complex detector arrangements and to process and extract the relevant information. The Geant4 code requires geometry libraries which are complemented with a specific class capable of describing irregular convex polyhedral shapes as the AGATA elementary shapes [2].

In this program, the ratios of Ge isotopes in the detector materials are 21%, 28%, 8%, 36% and 7% for $^{70}$Ge, $^{72}$Ge, $^{73}$Ge, $^{74}$Ge and $^{76}$Ge, respectively. The position resolutions of the detectors are about 5 mm. Simulated full-energy efficiency and peak-to-total ratio (P/T) of AGATA for 1 MeV photons are 43% and 58%, respectively. In all cases, the data is packed and smeared in the standard way and a 5 keV energy threshold is applied.

The neutrons can travel long distances and interact within AGATA detectors. Neutrons mainly interact with the germanium detector material via elastic or inelastic scatterings. The elastic scattering points can be determined and eliminated in the tracking process. Besides, the inelastic scattering of neutrons cause gamma-rays inside the detector materials (Ge) due to the fact the excited nuclei. These excited nuclei emit gamma-rays to fall down to their ground states. During this process, Ge nuclei recoil. Recoiling points [11] are the marks of the inelastic scattering points. In the case of scattering inside Ge material, the emerged gamma-rays are in some characteristic energies. The most intense peak which is considered in this work is 596 keV from $2^+$ to $0^+$ state in $^{74}$Ge isotope.

In the present simulations, 20000 neutrons and 20000 gamma-rays were simulated. Energies of the mono-energetic neutrons emitted from the source position were 1 MeV. Therefore, the 596 keV gamma-rays emerge inside the detectors. We have also performed the simulation of 596 keV gamma-rays from the source position. The aim of this work is to separate these different kind of gamma-rays which have same energy. The output of the simulation program that contains the spatial coordinates and energy of each interaction points (phase space) in the detectors was used as input of the Mgt (Mars Gamma-ray Tracking) gamma-ray tracking program [12]. In other words, the phase space data was transmitted simulation program to Mgt.

## 2.2 Gamma-ray Tracking

Two commonly used gamma-ray tracking algorithms exist; forward tracking [3] and backtracking [13]. Mgt program, which is used in this work, uses the forward tracking algorithm. The algorithm consists of two main stage. In the clustering stage, the interaction points are clustered according to the previously determined angular separation. This stage continues until all points are clustered. There might be clusters which have one element. These clusters could likely be belong to the elastic scattering points of neutrons and they are eliminated. In the cluster analysis stage, Compton scattering formula (Eq. 2.1) is used

$$E_{S1} = \frac{E_T}{1+(E_T/m_e c^2)(1-cos\theta)} \qquad (2.1)$$

where $E_T$ is the total energy of the initial gamma-rays, $\theta$ is the scattering angle and $E_{S1}$ is the energy of the scattered gamma-rays. Also, scattered energy can be obtained from the Eq. 2.2

$$E_{S2} = E_T - E_i \qquad (2.2)$$

where $E_i$ is the energy deposited in the i$^{th}$ scattering point. The difference between these two scattered energy can help calculation of figure-of-merit (FM) values of the clusters (Eq. 2.3)

$$FM = exp(-\frac{E_{S1}-E_{S2}}{\sigma_e^2}) \qquad (2.3)$$

where $\sigma_e^2$ is the uncertainty in scattering energy due to the uncertainty in the interaction coordinates resulting from finite position resolution (5 mm.) of the detectors. The analyses have been performed for all the clusters. After the analyses, the clusters are scored as "good", "bad" or "acceptable" according to the previously specified threshold value.

The clusters including Ge recoiling point are potentially bad clusters, whereas the clusters including only gamma-ray interaction points are good. This mark is the one criterion (FM) for neutron-gamma separation. In the gamma-ray tracking process, the energies deposited in the first interaction points of the radiations can be determined. In the case of neutrons, the first interaction points are generally recoiling energy points. The characteristic of the recoiling energies were investigated previously [2]. This mark is another criterion (efirst) for separation. The last criterion ($\Delta\theta$) for separation is scattering angle from the first interaction points. The angle is characteristically different for neutrons and gamma-rays [4]. In this study, these there criteria have been used in ANN for separation.

## 2.3 Artificial Neural Networks (ANN)

Artificial neural networks (ANN) [14] are mathematical models that mimic the human brain. They consist of several neurons which are processing units and they are connected each other via adaptive synaptic weights. By this synaptic connections, the neurons in the different layers communicate each other and the data is transmitted between them. In our calculations, we have used feed-forward ANN with three layers in order to make separation between neutrons and the gamma-rays. Three layer models with one hidden layer are recommended [15]. The first layer called input layer consist of four neurons, the hidden layer is composed of 50 neurons and the last one is the output layer with two neuron. The inputs were three criteria (FM, efirst and $\Delta\theta$) and the total energy of the cluster (eclust). The first and second outputs were 1, 0 for neutrons and 0, 1 for the gamma-rays. The used architecture of the ANN was 4-50-2 (Fig. 2.2) and the total numbers of adjustable weights were 300. The number of hidden layers depends on problem nature but in nearly all problems one hidden layer is sufficient [15]. No bias was used. The input neurons collect data from the outside and the output neurons give the results. The hidden neuron activation function was tangent hyperbolic (tanh = $(e^x-e^{-x})/(e^x+e^{-x})$) which is sigmoidlike function. For details of ANN, we refer the reader to [14].

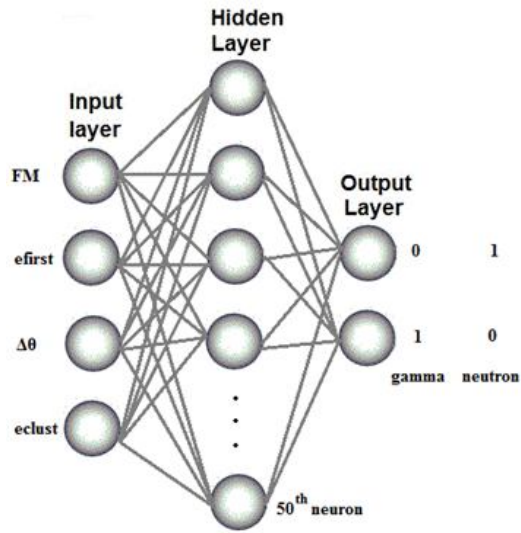

**Figure 2.2** The used ANN topology (4-50-2). The inputs were FM, efirst, Δθ and eclust. The outputs are 1 and 0 for neutrons, 0 and 1 for the gamma-rays.

The ANN method would be a perfect tool for separation of neutrons and the gamma-rays. The ANN processes are composed of two main stage, training and test stages. One has to train network by known data and then to feed the trained network with unknown data in order to obtain neural network outputs. The initial weights of the ANN are determined randomly by an algorithm. The training algorithm determines how the weights are changed [15]. In the training stage of this work, most popular training algorithm, back-propagation algorithm, with Levenberg-Marquardt [16,17] were used for the training of the ANN. In this algorithm, the net operates in two stages. In the first stage, a pattern is presented to the layer of the net input. Then, outcome activity flows through the net until the output layer generates the answer. Then, the output is obtained and compared with the desired output. The error function which measures the difference between wanted and actual neural network outputs was mean square error (MSE) given in Eq. (2.4). The MSE values are 0.08 and 0.1 for the training and test stages, respectively. In the training stage, 20000 data points were used. The test of the trained ANN was performed on the another data set including 20000 data points which have been never seen before by the network.

$$MSE = \frac{\sum_{k=1}^{r}\sum_{i=1}^{N}(y_{ki}-f_{ki})^2}{N} \qquad (2.4)$$

where $y_{ki}$ and $f_{ki}$ are neural network output and desired output, respectively, N is the number of training or test samples, whicever applies. For detailed information about using ANN in experimental nuclear physics, we refer to readers to the most recent works [15] and references therein.

### 3. Result and Discussions

In this study by using ANN, the separation power between 596 keV gamma-rays which are originated from source position and are from the inelastic neutron scatterings are increased. According to the previous work [4] which the criteria are determined and applied for separation, the rejection percentage of neutrons was about 69%. Besides, after applying the criteria as a gate, the lost percentage in the gamma-rays was 22% (Table 3.1). In this type of works of discrimination neutrons from gamma-rays, it is desired to reject all the neutrons and no gamma-rays. But using such a gate or any technique, a part of the gamma-rays are also rejected. The less rejection ratio in the gamma-rays gives better results. By the assist of the ANN method in this actual work, these ratios are increased from 69% to 88% for neutrons and decreased from 22% to 11% for the gamma-rays. ANN as a black box can identify the data whether it belongs to the neutrons or gamma-rays. This recognition ability of the ANN helps and makes easier solution of the problems.

**Table 3.1.** The rejection ratios (in percentage) and improvement factors for neutrons and the gamma-rays.

|  | Rejection Percentage | | |
|---|---|---|---|
|  | Using Gates | Using ANN | Improvement Factor |
| 1 MeV neutrons | 69% | 88% | 1.17 |
| 596 kev gamma-rays | 22% | 11% | 0.50 |

In Figs. 3.1 and 3.2, the gamma-ray spectra of the gamma-rays and neutrons are given, respectively. The spectra which forward gamma-ray tracking process were used called tracked spectra. As can be clearly seen in the Fig. 3.1, the lost in the gamma-rays is in the Compton continuum area by using ANN. The gamma-rays in this area belonging to the 596 keV photopeak are not desired. Therefore, such a lost in the gamma-rays are desired situation in this manner. In Fig. 3.2, it is seen that the neutrons are largely reduced by using ANN method. The 596 keV peak belonging to the inelastic scatterings of neutrons are dropped about its half level, whereas the peak of 596 keV gamma-rays remains nearly same level in Fig. 3.1. Additionally, other peaks and counts in Fig. 3.2 are highly reduced. In brief, this application save the intense

Improvement studies on neutron-gamma separation in HPGe detectors of the original 596 keV gamma-ray peak (Fig. 3.1) which is emerged from the gamma-rays emitted from the source position, whereas it elimiates neutrons which cause unwanted background source in the gamma-ray spectra.

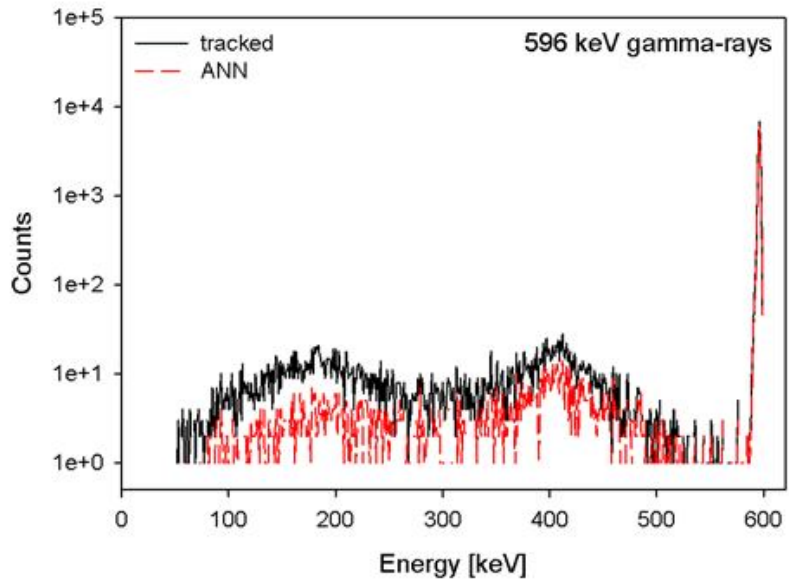

**Figure 3.1** The used ANN topology (4-50-2). The tracked energy spectrum of 596 keV gamma-rays (black line). Decreasing in counts after application of ANN method with three criteria (red dashed line). The most of the counts are rejected in Compton area.

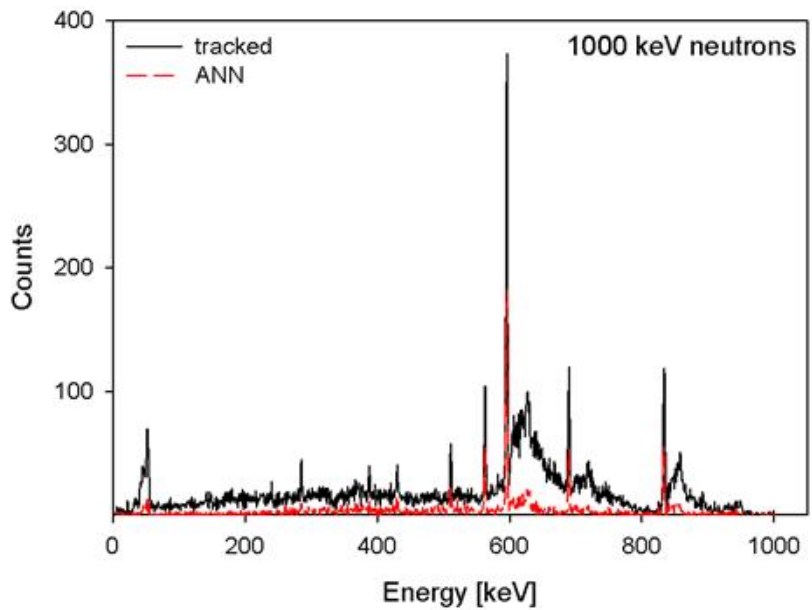

**Figure 3.2** The tracked energy spectrum of 1 MeV neutrons (black line). Decreasing in counts and particularly in 596 keV peak after application of ANN method with three criteria (red dashed line).

**Conclusions**

By using these three criteria and deposited total energy of the radiation, ANN method has been applied for separation of neutrons and the gamma-rays. In the starting stage of the work, only 596 keV gamma-rays and 1 MeV neutrons were considered interacting in the HPGe detectors. According to the preliminary results, the discrimination between neutron and the gamma-rays can be improved by using ANNs. The rejection ratio for neutrons is increased by a factor of 1.27. Besides the lost ratio for the gamma-rays is decreased by a factor of 0.50. In the case of realistic nuclear reactions, there are lots of neutrons and gamma-rays with different energy and multiplicity values. Therefore, it is planned to generalize the method for all neutron and the gamma-ray energies and the detectors.